\documentclass[useAMS,usenatbib,usegraphics]{mn2e}
\usepackage{graphicx,epsfig}  
\usepackage{amsmath,amssymb}
\usepackage{subfigure}
\usepackage{longtable}\usepackage{lscape}
\usepackage{times}
\def\ut#1{\mathop{\vtop{\ialign{##\crcr
     $\hfil\displaystyle{#1}\hfil$\crcr\noalign
     {\kern1pt\nointerlineskip}\hbox{$\hfil\sim\hfil$}\crcr
     \noalign{\kern1pt}}}}}

\def\undersymbol#1#2{\mathop{\vtop{\ialign{##\crcr
     $\hfil\displaystyle{#2}\hfil$\crcr\noalign
     {\kern1pt\nointerlineskip}\hbox{$\hfil#1\hfil$}\crcr
     \noalign{\kern1pt}}}}}

\newcommand{\integral}{{\it INTEGRAL}}
\newcommand{\swift}{{\it Swift}}
\newcommand{\xmm}{{\it XMM-Newton}}
\newcommand{\chandra}{{\it Chandra}}
\newcommand{\rxte}{{\it RXTE}}
\newcommand{\sou}{{IGR J17361--4441}}
\def\ergcms{erg cm$^{-2}$ s$^{-1}$ }
\def\ergs{erg s$^{-1}$}

\title[IGR J17361-4441: a possible planetary TDE]{The puzzling source \sou\ in NGC 6388: a possible planetary tidal disruption event
}
\author[M. Del Santo et al.]{M. Del Santo$^{1, 2}$, A.A. Nucita$^{3, 4}$, G. Lodato$^{5}$, L. Manni$^{3, 4}$, F. De Paolis$^{3, 4}$, J. Farihi$^{6}$,\and
 G. De Cesare$^{7, 2}$, A. Segreto$^{1}$ \\
 $^{1}$Istituto Nazionale di Astrofisica, IASF Palermo, Via U. La Malfa 153, 90146 Palermo, Italy\\
$^{2}$Istituto Nazionale di Astrofisica, IAPS, via Fosso del Cavaliere 100, 00133 Roma, Italy \\
$^{3}$Dipartimento di Matematica e Fisica {\it E. De Giorgi}, Universit\'a del Salento, Via per Arnesano, CP 193, I-73100, Lecce, Italy \\ 
$^{4}$INFN, Sez. di Lecce, via per Arnesano, CP 193, I-73100, Lecce, Italy\\ 
$^{5}$Dipartimento di Fisica, Universit\'a Degli Studi di Milano, Via Celoria 16, 20133 Milano, Italy\\
$^{6}$Department of Physics and Astronomy, University College London, London WC1E 6BT \\
$^{7}$ Istituto Nazionale di Astrofisica, IASF Bologna, via Piero Gobetti, 101, 40129 Bologna, Italy\\
}

\begin{document}

\date{Accepted 2014 July 16.  Received 2014 July 16; in original form 2014 March 21.}
\pagerange{\pageref{firstpage}--\pageref{lastpage}} \pubyear{2014}

\maketitle
\label{firstpage}

\begin{abstract}
On 2011 August 11,  \integral\ discovered the hard X-ray source \sou\ near the centre of the globular cluster NGC 6388.  
Follow up observations with \chandra\ showed the position of the transient was inconsistent with the cluster dynamical centre, 
and thus not related to its possible intermediate mass black hole.  
The source showed a peculiar hard spectrum ($\Gamma \approx 0.8$) and no evidence of QPOs, 
pulsations, type-I bursts, or radio emission.  
Based on its peak luminosity, \sou\ was classified as a very faint X-ray transient, and most likely a low-mass X-ray binary.
We re-analysed 200 days of \swift/XRT observations, covering the whole outburst of \sou\ and find a $ t^{-5/3}$ trend evident in the light curve, 
and a thermal emission component that does not evolve significantly with time.  
We investigate whether this source could be a tidal disruption event, and for certain assumptions find an accretion 
efficiency $\epsilon\approx 3.5\times10^{-4} ({M_{\rm Ch}}/M)$ consistent with a massive white dwarf, 
and a disrupted minor body mass $M_{\rm mb} \approx 1.9\times 10^{27}(M/{M_{\rm Ch}})$ g in the terrestrial-icy planet regime.
These numbers yield an inner disc temperature of the order k$T_{\rm in}\approx 0.04$ keV, 
consistent with the blackbody temperature of k$T_{\rm in}\approx 0.08$ keV estimated by spectral fitting.  
 Although the density of white dwarfs and the number of free-floating planets are uncertain, 
we estimate the rate of planetary tidal disruptions in NGC 6388 to be in the range $3 \times 10^{-6}$ to  $3 \times 10^{-4}  {\rm yr}^{-1}$.
Averaged over the Milky Way globular clusters, the upper limit value corresponds to  0.05 yr$^{-1}$, consistent with the observation
of a single event by \integral\ and \swift.
\end{abstract}

\begin{keywords}
accretion, accretion disc Ðwhite dwarfs Ð globular clusters: individual: NGC 6388 ÐX-rays: individual: IGR J17361Ð4441
\end{keywords}

\section{Introduction}\label{sec:intro}
The globular cluster (GC) NGC 6388 is considered one of the best candidates  
to harbour an intermediate mass black hole (hereafter IMBH; \citealt{baumgardt2005}) at its centre. Optical observations allowed a first estimate 
of the IMBH mass of $\simeq 5700$ M$_{\odot}$ \citep{lanzoni2007}. 
Under the hypothesis that the IMBH is accreting the surronding matter, it is natural to expect signatures in the X-ray band. 
\xmm\ and \chandra\  observations towards NGC 6388 showed that 
several {\it X}-ray sources exist in the GC centre
(see \citealt{nucita2008}, \citealt{cseh2010}) so that a unique identification of the putative IMBH 
is missing. 

Based on the correlation  between the X-ray and radio flux from black holes (\citealt{merloni2003}), 
\citet{maccarone2004} realized that radio observations may be very useful to pinpoint faint objects 
in GCs. 
 Following this suggestion, \citet{cseh2010} combined radio observations of the central 
region of NGC 6388  with the {\it Chandra} X-ray flux of the IMBH candidate, in order to put 
an upper  limit ($3\sigma$) on the BH mass of $\simeq 1500$ M$_{\odot}$. 
A similar result was obtained combining high spatial 
resolution and wide-field spectroscopy of more than 600 stars 
in the direction of NGC 6388 \citep{lanzoni2013}.

A general excitement started on 2011 August 11 when the IBIS telescope \citep{ubertini2003} on-board the {\it INTEGRAL} 
satellite \citep{winkler2003} identified  a new hard X-ray source, labeled \sou,  
close to the NGC 6388 centre \citep{gibaud2011}. The discovery of this transient 
opened the possibility that we were witnessing the awakening of the IMBH. 

In order to explore the nature of \sou\, several observing campaigns, in particular in the X and radio domains \citep{bozzo2011}, have been performed. 
The X-ray position obtained with \swift/XRT was consistent with the centre of NGC 6388  \citep{ferrigno2011}.
However, {\it Chandra} follow-up revealed that \sou\ was located $\sim 2.7 '' $  from the cluster 
centre and it was not consistent with the position of any  known X-ray source \citep{pooley2011}.
The transient source in NGC 6388 was also the target of two \xmm\ slew  observations (see for details \citealt{nucita2012}) 
which revealed a spectrum softer than that initially observed by XRT. 

Based on the X-ray luminosity of $6-9 \times 10^{35}$ \ergs\ estimated with XRT, \citet{wijnands2011} classified \sou\ 
as very-faint X-ray transient (VFXT; see \citealt{wijnands06} for a discussion about VFXTs).
Considering the location in a GC, these authors proposed the system being a neutron star (NS) low-mass X-ray binary (LMXB).
However, they noted that the observed photon index of $\Gamma \sim 0.6-1.0$ is atypical for LMXB transients in this luminosity range, where normally the photon index is around 1.6 -- 2.2 
 (e. g. \citealt{delsanto2007, degenaar2009}). 

After the discovery, \swift/XRT followed the evolution of the \sou\ outburst for $\simeq$ 200 days 
until 2011 November 5th when the source was no longer observable \citep{bozzo2012}.
We have reanalysed these XRT data (obtaining consistent results) and performed a deeper analysis of the light curve and spectral evolution.

We noted that a $\propto t^{-5/3}$ trend is clearly evident in the light curve and we started to investigate whether this source might be a tidal disruption event (TDE).

TDE have been observed over the years as X-ray \citep{komossa99, bloom11, cenko12, esquej07} and UV \citep{gezari2012, gezari2009} transients. 
In most cases, such outbursts follow from the disruption of a star by a supermassive black hole (SMBH), 
which results in a powerful super-Eddington flare with a duration of several weeks. 
Recently, it has also been shown that such TDEs can also produce jets, resulting also in radio emission \citep{zauderer11, levan11, bloom11}. 
In a few cases, the luminosity of the outburst appeared to imply that the disrupted object was a relatively low mass object, such as a planet or brown dwarf. 
This is the case, for example, of the giant X-Ray Flare in NGC 5905 \citep{li02}, or, more recently, of IGR J12580+0134, a flare discovered by \integral\ 
in 2011 and associated with NGC 4845 \citep{nikolajuk2013}. 

In both cases, the best fit to the light curve implied the tidal disruption of a giant planet by a SMBH.
 In one other case, \citet{campana11} had interpreted the GRB101225A as a Galactic TDE, 
 where an asteroid had been accreted by a NS. Such interpretation has, however, been ruled out by a redshift measurement of the source \citep{levan14}, 
 that demonstrates its extragalactic origin. 

Still, the sudden accretion of rocky objects by compact objects in the Galaxy is probably not uncommon. 
 For example, there is substantial evidence for tidal disruption of asteroids by white dwarfs (WDs).  
 To date, there are 30 published cases where debris is seen via thermal emission in the infrared  (e.g. \citealt{farihi10}), 
 and sometimes also gaseous metallic emission in the optical (e.g. \citealt{gansicke06}), 
 and via accreted metals in the stellar atmosphere. These circumstellar discs lie completely within the Roche limit of the stars, 
 approximately $1R_{\odot}$, and are thus consistent with tidally disrupted planetesimals  \citep{jura03}. 
 The planetary nature of the disc material is given both by the abundance ratios in the polluted stellar atmospheres  (e.g. \citealt{gansicke12}),
 and via infrared emission of evolved solids associated with planet formation \citep{jura09}.

Among these polluted WDs, there appears to be evidence in support of intense accretion episodes of planetary debris \citep{farihi13}.  
Briefly, trace metals in an H-rich WD provide instantaneous measures of accretion rates, 
which are typically in the range $10^5$ -- $10^8$ g s$^{-1}$ \citep{koester06}. 
On the other hand, atmospheric metals in a comparable He-rich WD give a historical average of accretion 
over time-scales up to $10^6$ yr \citep{farihi09}. Remarkably, while there are examples of stars with time-averaged accretion rates 
up to $10^{11}$ g s$^{-1}$, no instantaneous accretion rate above $10^9$ g s$^{-1}$ has yet been witnessed 
among many dozens of metal polluted WDs. Analysis of available data suggest that there are short-lived bursts ($t_{\rm high} < 10^3$ yr)  
of high-rate accretion in these polluted WD systems, such that the product of $t_{\rm high}\dot{M}_{\rm high}$ is around $10^{24}$ g, 
or 1/6000 the mass of Earth \citep{farihi12}.

The structure of the paper is as follows. In Section 2, we briefly discuss the \integral\ and \swift\ data analysis. 
In Section 3, we analyse the spectral evolution and the lightcurve of the system, demonstrating a remarkable $t^{-5/3}$ decline, typical of TDEs. 
In Section 4, we discuss the possible interpretation of the source: after ruling out a possible X-ray binary scenario, 
we concentrate on a TDE interpretation, showing that the parameter set that best describe the system indicates the disruption of a rocky object by a WD. 
We also discuss and rule out a possible interpretation as an extragalactic TDE in a background galaxy. 
In Section 5 we draw our conclusions.

\begin{figure*}
\centering
\includegraphics[height=8cm]{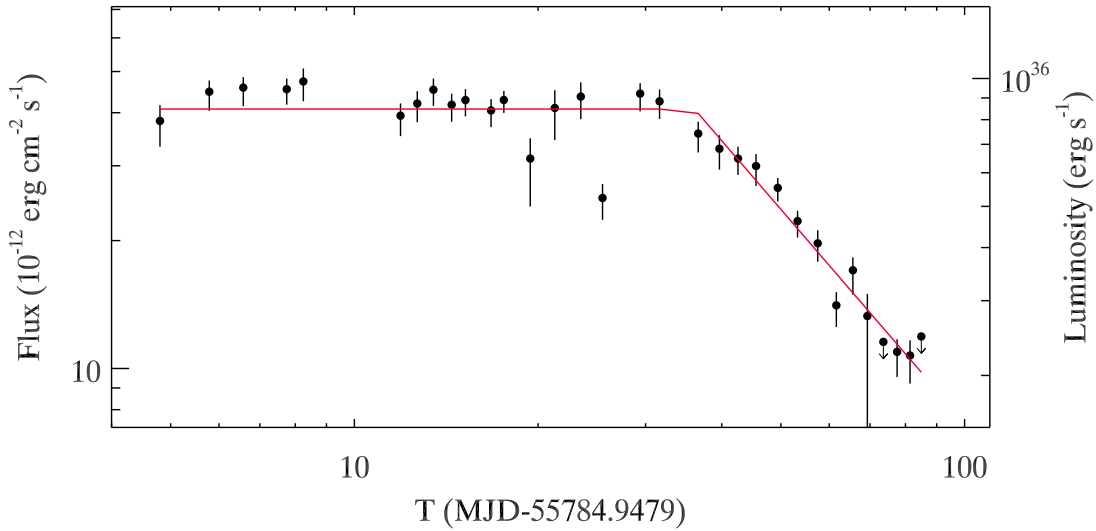}
\caption{The 1--10 keV \swift/XRT light curve corrected for the absorption.
The light curve has been fitted with a broken power law, i. e. a constant plus an exponential which
results in a $\propto t^{-5/3}$ law. }
\label{fig:xrt_flux}
\end{figure*}

\section{Observation and data analysis}
\subsection{Swift/XRT}
We have analysed the \swift/XRT observations performed on the target \sou\ (Id. numbers from
00032072001 to 00032072038; see also Bozzo et al. 2012) for a total exposure time of $\simeq$ 105 ks. 
\swift\  data have been analysed by using the standard procedures described
in \citet{burrows2005} and with the latest calibration files\footnote{${\rm http://heasarc.nasa.gov/docs7swift/analysis/.}$}. 
We processed the XRT data with the {\sc xrtpipeline} (v.0.12.6) task and
we applied standard screening criteria by using ftools (Heasoft v.6.13.0). 
The source spectra have been extracted from a circular region centred on the target nominal coordinates (with radius
of $\simeq 40''$) by using the {\sc xselect} routine. The background spectra were obtained from circular regions far enough
from the \sou\ position. When necessary, we corrected observations affected by pile-up (i.e. with source count rates
above $ \simeq 0.5$ ct s$^{-1}$). We then used the {\sc xrtmkarf} task to create the ancillary response files. 
Of the 38 XRT spectra, we avoided to use the last six because of the low statistics.
Each of the 32 spectrum has been rebinned in order to have at minimum 25 counts per channel with {\sc grppha}. 
All fits have been performed with {\sc xspec}  \citep{arnaud07} and errors calculated at 90$\%$ confidence level.

\subsection{INTEGRAL/IBIS}
We have analysed all data collected by the low energy IBIS detector, i.e. ISGRI \citep{lebrun03},
when \sou\ was present in the total IBIS  field of view (14.5$^{\circ} \times$14.5$^{\circ}$).
Observations have been performed in the period 2011 August 11 -- October 22, i.e. \integral\ revolutions 1078--1101 
for a total of 454 science windows (SCW) and $\simeq$1 Ms time exposure. 
Data analysis was carried out with the Off-line Scientific Analysis (OSA) v. 10.0
distributed by the ISDC \citep{cou03}.
We processed data with the {\sc ibis\_science\_analysis} task and extracted images in three energy bands: 18--40, 40--80, 80--150 keV.
We did not get any detection at SCW level ($\sim$ 2 ks), so that we mosaicked the 454 images in groups of 38, 39, 61, 63, 130 and 123.
A significant signal to noise (above 5 $\sigma$) has been obtained only in the first energy band (18--40 keV).
The errors on the IBIS/ISGRI count rate have been obtained via the significance (signal-to-noise ratio) in output from the OSA software.
We extracted an averaged ISGRI spectrum (53 ks) corresponding to the flat part of the XRT light curve (see Secion \ref{sec:spec}) 
by using the OSA task {\sc mosaic\_spec} (useful for faint sources).

\subsection{Swift/BAT}
The \swift/BAT survey data collected from 2004 November up to  to end of 2013 retrieved from the HEASARC public archive were processed using 
{\sc batimager} \citep{segreto10}. This code is dedicated to the processing of coded mask instrument data. 
We found that \sou\ is detected in only one time interval, since MJD $\sim$ 55770 up to MJD $\sim$ 55828, with a statistical significance
of 15 standard deviations in the 20--85 keV all-sky map.
Lightcurve has been extracted  with 5 days temporal bins.

\begin{figure}
\centering
\includegraphics[height=8cm]{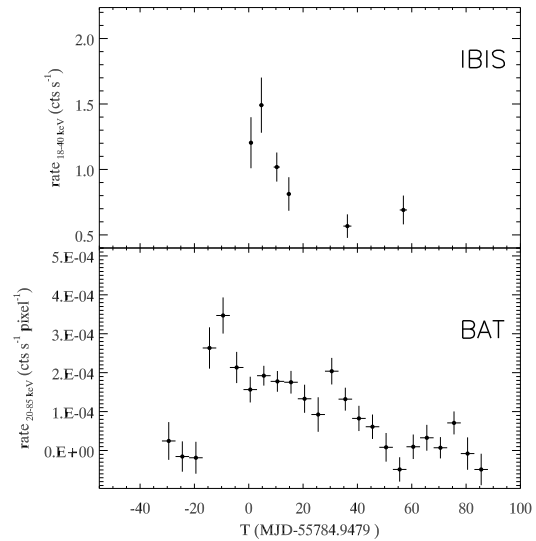}
\caption{IBIS (18--40 keV) and BAT (20--85 keV) light curves. The BAT detection occurred 14 days before the IBIS one.}
\label{fig:bat_lc}
\end{figure}

\section{Results}\label{sec:results}

\subsection{Temporal analysis}
We have estimated the  $1-10$ keV unabsorbed fluxes by fitting the 32  \swift/XRT spectra 
of \sou\ with a simple power-law. Plotting the flux evolution, we noted that it could be well represented by a broken power-law.
Thus, we fit the light-curve with a constant plus an exponential law (see Fig. \ref{fig:xrt_flux}).
It results a plateau flux $F_{\rm max}$= ($4.1 \pm 0.1) \times 10^{-11}$ \ergcms\ and a knee time of $t_{\rm k}$= $36 \pm 1$ 
days from the \integral\ trigger  MJD$_{\rm IBIS}= 55784.9479$, i.e. roughly 5 days before the first XRT follow-up observation.
Then, it was very interesting to observe that after the knee the data were  well fit with  a $\propto t^{-5/3}$ law ($\chi_{\nu}^{2}$= 1.14 with 30 d.o.f.).

In Fig. \ref{fig:bat_lc}, the \integral/IBIS (up) and \swift/BAT (bottom) hard X-ray light curves  are shown.
Thanks to BAT, we found that the \sou\ outburst was started (in hard X-rays) roughly 14 days (MJD$_{\rm BAT}$ = 55770.4480)
before the \integral\ trigger, thus implying a total event (observed) duration of 99 days, when the last XRT point (MJD$_{\rm stop}$ = 55869.8738 day) is considered as the end.

\subsection{Spectral analysis}\label{sec:spec}
\begin{figure}
\centering
\includegraphics[height=8cm]{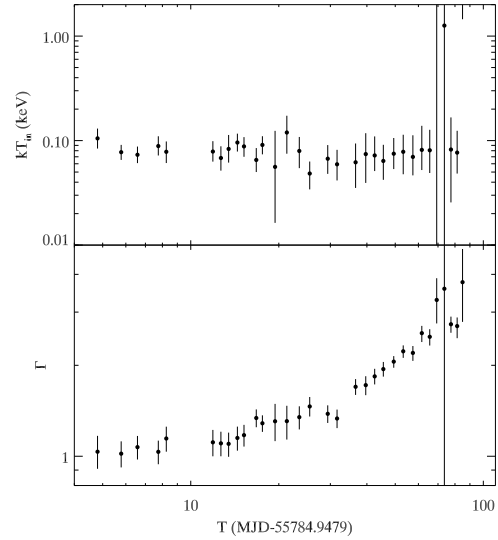}
\caption{\swift/XRT spectral evolution of the 0.3--10 keV spectra fit with a disc blackbody component (k$T_{{\rm in}}$) plus a simple power-law ($\Gamma$).}
\label{fig:xrt_fit}
\end{figure}

Extending the XRT spectral energy range down to 0.3 keV, an additional disc blackbody component to the power-law is requested by the data. 
Then, the 32 XRT spectra of \sou\ have been fit with an absorbed power-law plus {\sc diskbb} (in XSPEC).
The column density is constant at a mean value of $\simeq 0.8 \times 10^{22}$ cm$^{-2}$ (also consistent with  \citealt{bozzo2011}), so that 
it has been fixed in our fit procedure.
The evolutions of the inner disc temperature (k$T_{\rm in}$) and power-law slope ($\Gamma$) are shown in Fig. \ref {fig:xrt_fit}.
It is interesting to note that in spite of the variation of the power-law component,
i. e. an abrupt change of the power-law slope at roughly $t_{\rm n} \simeq$ 30 days from the \integral\ trigger
is observed, the disc temperature results constant at a value around 0.08 keV (Fig. \ref {fig:xrt_fit}).

To obtain the hard X-ray spectrum up to $\simeq$100 keV, we averaged the IBIS/ISGRI data corresponding to the first four bins of the light curve (Fig. \ref{fig:bat_lc}, {\it top}).
These data are simultaneous to the plateau period in the XRT light-curve, when \sou\ did not show any spectral variation (Fig. \ref{fig:xrt_fit}).
Then, we have combined the IBIS/ISGRI spectrum with one of the brighter XRT spectra and fit
the broad-band spectrum with an absorbed cut-off power-law ({\sc cutoffpl}) plus {\sc diskbb}, 
obtaining parameters (see Table \ref{tab:broad_fit}) consistent with those reported in \citet{bozzo2011}.

In addition, we fit this spectrum with a physical model by using a thermal Comptonization model ({\sc comptt} in XSPEC) in place of the cutoff power law
obtaining an electron population temperature of roughly 10 keV, an inner disc radius $R_{in} \simeq 12600$ km (from the normalization of the {\sc diskbb} model; \citealt{mitsuda1984})
 and a bolometric (0.1--100 keV) flux of about 2$\times 10^{-9}$ \ergcms\ (see Tab. \ref{tab:broad_fit}).
Assuming a distance of 13.2 kpc \citep{dalessandro2008}, we have estimated the bolometric peak luminosity as $L_{\rm bol}\simeq 3.5 \times 10^{37}$ erg s$^{-1}$. 

It is known that  in accreting compact objects $L_{\rm X}$ corresponds (as first approximation)  to the accretion luminosity $L_{\rm acc} = \epsilon \,  \dot{M} \,  c^2 $. 
Thus, a lower limit to the accreted mass  on to the central object can be evaluated. 

We have integrated the bolometric luminosity over the time for the whole outburst duration and obtained $M_{\rm acc} \simeq 3.4 \times10^{23}\,\epsilon^{-1}\, \rm g$.

\begin{figure}
\centering
\includegraphics[height=6cm, angle=-90]{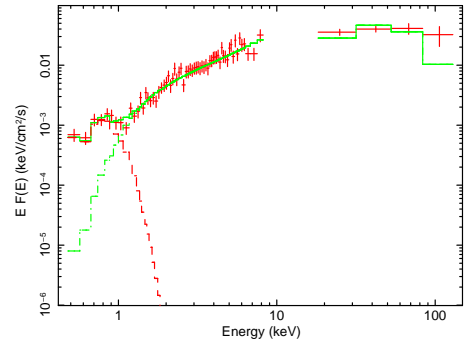}
\caption{XRT+IBIS spectrum fit with an absorbed disc blackbody plus Comptonization component.}
\label{fig:ibis_spec}
\end{figure}

\begin{table*}
\renewcommand{\arraystretch}{1.5} 
\begin{center}
\caption{XRT plus IBIS broad-band spectrum fit with both an empirical cut-off power law ({\sc cutoffpl})
with $\Gamma$ as the power law photon index and $E_{\rm c}$ the energy cut-off, and a physical thermal Comptonization model ({\sc comptt})
with  k$T_{\rm e}$  and $\tau$ as the temperature of  Comptonizing electrons and the Thomson depth, respectively. 
A multicolour disc blackbody with (k$T_{\rm in}$) temperature  is also requested by the low-energy data, as well as the galactic absorption ($N_{\rm H}$).}       
\label{tab:broad_fit}      
\begin{tabular}{l c c c c c c c c} 
   
\hline\hline
Model    &     $N_{\rm H}$ &    k$T_{\rm in}$     & $E_{\rm c}$        &  $\Gamma$  &  k$T_{e}$           &  $\tau$         & $\chi^2_{\nu}$(dof)  &  $F_{\rm bol}$ \\ 
             &          $ 10^{22} $ [cm$^{-2}$]      &            [keV]       &    [keV]                 &                       &   [keV]                &                        &                                     &     \ergcms    \\
\hline                    

{\sc diskbb+cutoffpl}   &   $0.7 \pm 0.1$       & $0.089^{+0.004}_{-0.005}$           & $ 41^{+18}_{-12}$        &      $0.8 \pm 0.1$  &         $-$        &   $-$     &    $0.80(62)$    &  1.0$\times 10^{-9}$    \\
{\sc diskbb+comptt}   &  $0.7 \pm 0.1$         & $0.087^{+0.004}_{-0.006}$       &  $-$ &   $-$    &        $12 \pm 2$          &  $8 \pm 1$&  $1.01(62)$          &     1.7$\times 10^{-9}$\\

\hline\hline                                
\end{tabular}

\end{center}
\end{table*}

\section{Discussion: what kind of source can it be?}

\subsection{X-ray binary scenario}\label{sec:binary}

Based on the  peak luminosity of $L_{\rm 2-10 keV}$= $8.5 \times 10^{35}$ \ergs, 
\sou\ has been classified as VFXT \citep{wijnands2011}.
This is a non-homogeneous class of sources characterized by a peak luminosity (in 2--10 keV) within the range $10^{34-36}$ \ergs\ \citep{wijnands06}.
Very faint transients in GC are most likely LMXB with NS or BH as accreting compact object.

As discussed in details in   \citet{kulkarni93}, \citet{portegies00} and \citet{strader12} 
one expects that even if hundreds of stellar-mass BHs form in a typical GC, 
only a few of them survive since the others are likely
ejected through dynamical interactions.

According to \citet{kalogera2004}, they are likely to have extremely low duty cycles (below about $10^{-3}$),
consistent with the absence of such an object in Galactic GC.
Observational support for such picture was provided by the lack of X-ray binaries with BH in Milky Way GC
until 2011, when \citet{strader12} proposed that two radio sources in the GC M22 are stellar-mass BH.

The possibility that \sou\ is a binary system with a BH  can be ruled out because of the
low radio-to-X-ray flux ratio (see \citealt{chomiuk13, bozzo2011}). 
\citet{ferrigno11} reported on the ATCA non-detection of \sou; they inferred an upper limit of 19 $\mu Jy$ at 9 GHz
which translates into a luminosity at 8.5 GHz of $4 \times 10^{28}$ \ergs. 
We note that this value is at least a factor of 100 fainter than the typical radio luminosity observed 
in BH binaries with L$_{\rm X} \simeq 8 \times 10^{35}$ \ergs (see Fig. 3 in \citealt{migliari06}).
Moreover, BH transients in the hard state usually show quasi-periodic oscillation (see \citealt{mcclintock06, homan05}).
These features were also observed in the very faint  system Swift J1357.2--0933 \citep{casares2011, armas2014}.
However, the power density spectrum constructed after accumulating \rxte/PCA  lightcurves of \sou\ did not show the presence of any coherent signal (against white noise)
in the frequency range $10^{-3}$ Hz - $10^{3}$ Hz  at 3 $\sigma$ confidence level  (see Fig. 2 in \citealt{bozzo2011}).

The other possibility is that \sou\ is a LMXB with a NS accreting at very low $\dot{M}$,
either  with low magnetic field (possibly showing suddenly type-I X-ray bursts, i.e. burst-only sources; \citealt{cornelisse2004}, \citealt{delsanto2007}, \citealt{campana2009})
or with high magnetic field, i. e. an accreting millisecond X-ray pulsars (AMSP; \citealt{wijnands08, wij08}).

However, the hard X-ray spectrum ($\Gamma \simeq 0.8$; Tab. \ref{tab:broad_fit}) observed at the beginning of the outburst of \sou\  is not observed in VFXT with low magnetic field NS.
Usually, these objects show $\Gamma \sim 1.5 - 2.2$ in this luminosity range (see \citealt{degenaar2010} and references therein) and blackbody temperatures 
(when present) consistently higher than that observed in \sou\ (i. e. 0.3--0.5 keV; see \citealt{armas2013}).
Moreover, the  soft spectra observed at the end of the outburst are typical of the soft spectral state,
which is observed in NS burster sources only when in high accretion rate regime (i.e. luminosities; \citealt{barret00}).

The possibility that the system might harbor an AMSP was proposed at the onset of the \swift\ observations,
hypothesis which was immediately in contrast with the lack of pulsations in the PCA data \citep{bozzo2011}.
Even considering that pulsation in LMXBs  might be transient as observed in Aql X--1 \citep{casella2008},
we note that AMSPs have similar power law slopes than the non-pulsating NS and blackbody temperatures even higher, i. e. up to 1 keV (see \citealt{patruno12, papitto10}).
Moreover they usually do not show variations of the spectral shape at different flux levels (see \citealt{falanga12} and references therein).

We conclude that the hypothesis either of low magnetic (burster) or high magnetic (AMSP) NS for the compact object of  \sou\  is unlikely.
In addition, this is supported by the fact that neutron star transients have been observed to show (usually) quiescent luminosities 
of about  $10^{32-33}$ \ergs (see \citealt{rea2011}, and references therein), i. e. higher than the upper limit of few $10^{31}$ \ergs\ derived for \sou.

\subsection{Hints for a TDE?}
\begin{figure}
\centering
\includegraphics[height=4cm]{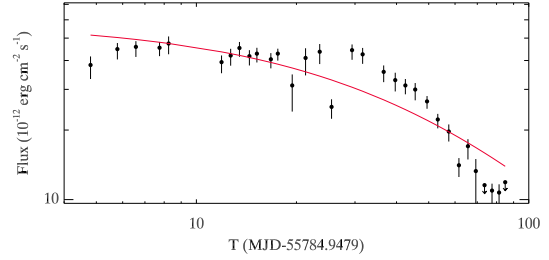}
\caption{XRT light curve fit with the model described in Eq. (\ref{eq:mdot})}
\label{fig:fit_nogood}
\end{figure}

Based on the considerations reported in Section \ref{sec:binary}, we suggest \sou\ being a TDE.
There are two empirical evidences that point in the direction of interpreting this flare of \sou\ as a TDE. 
First and foremost, the apparent decline of the light curve as $t^{-5/3}$, that is typical of such events, even though we can fit the expected decline only 
to the late light curve, while the early light curve shows a plateau which is generally not expected in TDEs. 
We tried to fit the light curve with a pure $t^{-5/3}$ type law and we did not obtain a good fit ($\chi^{2}_{\nu}$=2.3, 30 d.o.f.; see Fig. \ref{fig:fit_nogood} ). 

Secondly, several TDEs produced by the disruption of stars by a SMBH appear to be characterized by a thermal emission component 
that does not evolve significantly with time, as in the present case. Such component has been observed in TDEs discovered 
in the optical \citep{gezari2012,chornock13,vanvelzen11} and in X-rays \citep{donato2013,maksym13}. 
A slow evolution of the thermal component is not expected from theoretical models of TDE, 
that imply a cooling thermal component \citep{LR11}, unless the fallback rate is strongly super-Eddington \citep{loeb97}, 
or unless some reprocessing of the high-energy emission by an opaque shroud of debris is invoked (\citealt{guillochon13},
which would only be able to explain the events observed in the optical). 

In a TDE, the accreted mass is half of the mass of the disrupted object. According to the numbers quoted above (end of Sec. 3), 
we thus estimate that the disrupted minor body would have a mass of $M_{\rm mb} \approx 7 \times 10^{23} \epsilon^{-1}$ g. 
Even if the efficiency parameter is low, if the event is caused by accretion on to a compact object it would put the minor body 
in the asteroid -- terrestrial planet mass range, and would thus most likely be of rocky composition.

The standard theory of TDEs \citep{rees88,phinney89,evans89,LKP09} predicts that the debris of the disrupted object would return to pericentre at a rate given by:
\begin{equation}
\dot{M}_{\rm mb} = \dot{M}_{\rm p} \left(\frac{t+t_{\rm min}}{t_{\rm min}}\right)^{-5/3},
\label{eq:mdot}
\end{equation}
where $t$ is the time since the beginning of the flare and $t_{\rm min}$ is the return time of the most bound debris, given by
\begin{equation}
   t_{\rm min} = \displaystyle\frac{\pi}{2^{1/2}} \left(\frac{r_{\rm t}} {R_{\rm mb}}\right)^{3/2} \sqrt{\frac{r_{\rm t}^3}{G M}},
   \label{eq:tmin}
\end{equation}
where $M$ is the mass of the compact object, $R_{\rm mb}$ is the radius of the minor body and $r_{\rm t}$ is the tidal radius. 
Note the strong dependence of $t_{\rm min}$ on the tidal radius \citep{sari10, stone13}.
For an object with a density $\rho= 3$ g cm$^{-3}$ (appropriate for rocky bodies), the tidal radius is \citep{davidsson99,jura03} 
\begin{equation}
r_{\rm t} = C\left(\frac{3M}{4\pi\rho}\right)^{1/3}\approx 1.8R_{\odot}C\left(\frac{M}{M_{\rm Ch}}\right)^{1/3}\left(\frac{\rho}{3\mbox{g/cm}^3}\right)^{-1/3},
\label{eq:tidal}
\end{equation}
where  $C$ is a numerical factor dependent on the structure of the body and $M_{\rm Ch}$ is the Chandrasekhar mass. For a rocky object $C=1.26$. Inserting 
equation (\ref{eq:tidal}) into equation (\ref{eq:tmin}), we obtain
\begin{eqnarray}
   t_{\rm min} & = & \displaystyle\frac{C^3\pi}{2^{1/2}} \left(\frac{M} {M_{\rm mb}}\right)^{1/2} \sqrt{\frac{3}{4\pi G \rho}} \\
                      &  \nonumber \approx  & 1.4\times 10^4 \left(\frac{C}{2}\right)^3\left(\frac{M}{M_{\rm Ch}}\right)^{1/2}\epsilon^{1/2}\mbox{d},
   \label{eq:tmin2}
\end{eqnarray}
where for the mass of the minor body $M_{\rm mb}$ we have used $M_{\rm mb}=2M_{\rm acc}\approx 7\times 10^{23} \epsilon^{-1}$. 
The peak accretion rate is given by:
\begin{equation}
\dot{M}_{\rm p}=\frac{M_{\rm mb}}{3t_{\rm min}}.
\end{equation}
We can empirically obtain an estimate of the fallback time $t_{\rm min}$ by the following argument: if the light curve follows exactly a $t^{-5/3}$ decline law, 
then the total energy emitted, the peak luminosity and the fallback time are related by
\begin{equation}
t_{\rm min}=\frac{2E}{3L_{\rm p}}\approx 68\mbox{d},
\end{equation}
where in the last equality we have used the measured peak luminosity and integrated energy of the event. Note that a fit of the lightcurve with a
 function of the form shown in equation (\ref{eq:mdot}) would not provide a good fit (since it does not reproduce well the initial plateau, as mentioned above) 
 but would return a best fit value for $t_{\rm min}$ in the range of 70 days, consistent with the simple estimate above. By imposing that the $t_{\rm min}$ 
 estimated from the observations matches the one derived from tidal disruption theory, we obtain an estimate of the accretion efficiency
\begin{equation}
\epsilon\approx 3.5\times10^{-4}\left(\frac{M_{\rm Ch}}{M}\right),
\end{equation} 
which is consistent with the expected accretion efficiency of a WD close to the Chandrasekhar limit. 
With these estimate for the efficiency, we can obtain an estimate for the mass of the disrupted minor body and for the peak accretion rate as
\begin{equation}
M_{\rm mb} \approx 1.9\times10^{27} \left(\frac{M}{M_{\rm Ch}}\right)~\mbox{g},
\end{equation}
\begin{equation}
\dot{M}_{\rm p} \approx 3\times10^{20} \left(\frac{M}{M_{\rm Ch}}\right)~\mbox{g/s}.
\end{equation}
We thus conclude that the disrupted object mass is of the order of a third Earth mass, which is in the terrestrial-icy planetary regime. With such numbers, the inner disc temperature at peak, close to the WD surface, is of the order of k$T_{\rm in}\approx 0.04$ keV, consistent with the blackbody temperature estimated by spectral fitting (see Section 3.2). 

A critical question to be addressed is whether the disrupted material, that would initially orbit relatively far from the WD surface, would be able to accrete fast enough so that the accretion rate on to the WD matches the fallback rate given in equation (\ref{eq:mdot}). For this to happen, we need to require that the viscous time in the disc is much smaller than the fallback time $t_{\rm min}$. Two recent studies have addressed this question. \citet{metzger12} argue that most of the accretion occurs at a late stage after the disc formation, and estimate viscous times of the order of thousands of years. In this case, the WD would accrete at a rate much smaller than the fallback rate and the disc would shine mostly in the infrared. Conversely, \citet{bear13} conclude that the energy liberated during the formation of the disc is high enough to lead to a rapid accretion of the material, which would lead to an initial spike of intense accretion at rates comparable to the fallback rate.
 In this case, we do expect to see the system to emit in X-rays. Thus in order to explain this source as the tidal disruption of a terrestrial planet by a WD, we need to invoke the mechanism suggested by \citet{bear13}. 

The hard X-ray emission is interpreted to originate from inverse-Compton scattering of lower energy photons (kT$_{in} \simeq 0.09$ keV)
by high energy electrons in a hot corona (kT$_{e} \simeq 12$ keV; Tab. \ref{tab:broad_fit}) forming around the accretion flow (see also \citealt{nikolajuk2013}).
This is the same mechanism which is thought to occur in all compact objects accreting matter via an accretion disc.
In TDE, we do not have detailed models for the formation and evolution of the corona and the associated time-scales. 
On the other hand, hard X-ray emission is not uncommon in  TDE candidates \citep{nikolajuk2013, cenko12, burrows2011}, but
 it is difficult to make a comparison with our \sou\ event since they are all thought to have occurred on a SMBH.

\subsection{Estimating the rate of planetary tidal disruption}

In this section we estimate the rate of tidal disruption of planets by WD in the core of a GC. It should be noted 
that these estimates rely on a number of assumptions that are either poorly constrained observationally or poorly understood theoretically, so they should be treated with caution.

Numerical simulations have demonstrated that dissolution of planetary systems occurs on timescales that are shorter than the age 
of a typical open or globular clusters \citep{spurzem2008}. We thus expect that the disruption does not follow from the deflection of a planet that is initially bound to the WD.

However the interacting object may be a free-floating planet (FFP), resulting from the dissolution of planetary systems.
The retention of a population of FFP in GC is not obvious. Simple estimates based on equipartition arguments would imply that a planetary 
population would acquire an extremely large velocity dispersion, well in excess of the escape speed from the cluster, and would thus evaporate within a 
 few relaxation times (note that the relaxation time for NGC 6388 is $\approx 5.2\times10^7$ yrs at the core and $\approx 8\times10^8$ yrs at the half-mass radius; 
 \citealt{harris96}). However, numerical simulations \citep{fregeau2002, hurley2002} show that a significant population of FFP is retained in relaxed systems. 
 A possible explanation for this behaviour is due to a lack of equipartition. Indeed, again through numerical simulations including a mass spectrum, \citet{trenti2013} 
 have shown that, as a consequence of Spitzer instability, equipartition does not occur even for relaxed
stellar systems, and that the velocity dispersion depends very weakly on mass, such that $\sigma\propto m^{-0.08}$. In our particular case, with a
 mass of a third Earth mass and an average stellar mass of $0.1M_{\odot}$, we estimate that $\sigma_{\rm pl} \approx 2.5 \sigma_{\rm star}$. 

The rate of tidal encounters can be simply evaluated \citep{binney2008, soker2001}:
\begin{equation}
\dot{N}_{\rm TE} \simeq \frac{4\pi r_{\rm c}^3}{3}{N}_{\rm FFP} {N}_{\rm WD} \Sigma \sigma_{\rm pl}, 
\label{eq:rate}
\end{equation}
where ${N}_{\rm FFP}$ is the density of FFPs, ${N}_{\rm WD}$ is the density of WDs, $r_{\rm c}$ is the core radius of the cluster and $\Sigma$ is the cross-section of the interaction,
\begin{equation}
\Sigma = \pi {r_{\rm t}}^{2} \left(1+ \frac { 2GM} { \sigma_{\rm pl}^2 {r_{\rm t}} }\right),
\label{eq:sigma}
\end{equation}
where one has to take into account the effect of gravitational focusing, which is dominant in this case. 
Note that in equation \ref{eq:rate} we assumed that the main contribution comes from the inner part of the GC, within the core radius ($r_{c}$). 

 In order to estimate the rate of planetary disruptions, we need to know the number density of FFP and WD in a typical  
GC, both of which are very uncertain. Here, we provide a pessimistic estimate of the rate of disruptions by considering the lowest values for both parameters, 
and an optimistic estimate considering the highest values. For the WD number density we assume either a low value of $\approx 10^{4} \rm{pc}^{-3}$ \citep{raskin2009}, 
and a high value - more appropriate for collapsed GC, such as NGC 6388 - of $10^{5} {\rm pc}^{-3}$, in agreement with simulations performed by \citet{ivanova2006}. 
Estimating the number density of FFP of rocky composition is even more difficult. It has been estimated that the density of FFP in GC
could exceed ${N}_{\rm FFP}= 10^{6}$ pc$^{-3}$ \citep{soker2001}, i.e. approximatively one per star in NGC 6388 considering the star density  reported in \citet{pryor1993}. 
Indeed, {\it N}-body simulations  suggested that FFPs could evolve to the current epoch with a population that exceeds the stellar population at the 
cluster center by a factor of $\sim 100$ \citep{fregeau2002, hurley2002}. This is also supported by N-body simulation of planet formation performed by \citet{ida2003}. 
We thus assume a low value for ${N}_{\rm FFP}= 10^{7}$ pc$^{-3}$ (i.e. ten per star) 
and a high value of $10^{8}$ pc$^{-3}$ (i.e. 100 per star).

We take $r_{\rm c}=0.5$ pc for the GC core radius size \citep{lanzoni2007} and $\sigma_{\rm star}\approx 13$ km/s \citep{lanzoni2013}. 
The tidal radius $r_{\rm t}\approx 1.5 \times 10^{11}$ cm is estimated from Eq. (\ref{eq:tidal}). We thus obtain a pessimistic estimate of the rate of TDE as:
\begin{equation}
\dot{N}_{\rm TE} \simeq 3.3 \times 10^{-6}\mbox{yr}^{-1}, 
\label{eq:rate2}
\end{equation}
(where we have assumed $N_{\rm WD}=10^4$pc$^{-3}$ and ${N}_{\rm FFP}= 10^{7}$ pc$^{-3}$) and an optimistic estimate of
\begin{equation}
\dot{N}_{\rm TE} \simeq 3.3 \times10^{-4}  \mbox{yr}^{-1},
\label{eq:rate3}
\end{equation}
(where we have assumed $N_{\rm WD}=10^5$pc$^{-3}$ and ${N}_{\rm FFP}= 10^{8}$ pc$^{-3}$)

In the {\it HST}/ACS sources catalogue of NGC 6388 used by \citet{lanzoni2007} 
there are 31 point sources  within the \chandra\ error box of \sou.
No one is a WD (Lanzoni priv. comm.). A more detailed analysis of the data available in the HST archive will be presented in a forthcoming paper.
However, it is worth to note that only very bright WDs can be observable by  {\it HST} in the redden core of NGC 6388 .

\subsection{Extra-galactic TDE}

We considered the possibility that \sou\ has an extragalactic origin, i.e. it is due by a sudden increase in the  accretion rate
on an SMBH at the centre of an otherwise quiescent galaxy (see \citealt{komossa2004, burrows2011, saxton2012, nikolajuk2013}). 

First of all, we have estimated the number of background AGNs expected towards the target within a circle of radius $\simeq 0.6''$
through the log N--log S diagram \citep{hasinger2005}  and by using
the \chandra\ flux upper limit $\simeq 4.8\times 10^{-16}$  \ergcms  \citep{pooley2011}.
Assuming a power-law model with $\Gamma=1.7$ and an absorption column density $N_{\rm H}=3 \times 10^{21}\, {\rm cm^{-2}}$ \citep{dickey1990}, 
we have obtained an unabsorbed flux in the 0.5--2 keV band of $ 2.9\times\,10^{-16} $ \ergcms\ 
(by using webPIMMS v3.9\footnote{{\tt http://heasarc.gsfc.nasa.gov/Tools/w3pimms.html}}). 
The expected number of background AGNs is then $\simeq 1 \times 10^{-4}$.  Thus, the probability of the existence of a background AGN is unlikely although not negligible.

On the other hand, we have estimated the likelihood that the event is an extragalactic TDE occurring in a background galaxy, either with an AGN activity or not.
A decay time-scale of $\approx 70$ days is in agreement with the expected time-scale for the disruption of a solar type star 
by a SMBH. In particular we can reproduce the observed inner disc temperature at peak of k$T_{\rm in}\approx 0.08$ keV 
and a decay time-scale of 68 days if the disrupted object has a mass $M_{\star}=0.48M_{\odot}$ and the SMBH has a mass of $M\approx 5.8\times10^6M_{\odot}$. 

The expected observed bolometric flux in this case is given by:
\begin{equation}
F_{\rm bol}=\frac{1}{4\pi D^2}\frac{\epsilon M_{\star}c^2}{3t_{\rm min}},
\end{equation}
where $M_{\star}$ is the mass of the disrupted star. Now, assuming a SMBH accretion efficiency of $\epsilon=0.1$ and $M_{\star}=0.48M_{\odot}$,  
we can estimate the distance of the source based on the observed flux $F_{\rm bol}\approx 2 \times 10^{-9}$ \ergcms
 and the observed decay timescale $t_{\rm min}\approx 68$ days. We then obtained $D\approx 160$ Mpc, corresponding to a redshift of $z\approx 0.04$. 
 
 We can then use the \citet{marconi2003} $L_{\rm K,bul}$--M$_{\rm BH}$ relation to obtain the K-band luminosity of the putative host galaxy for the assumed BH mass of $\sim 5.8\times10^6M_{\odot}$. At a distance of $\sim$ 150 Mpc the resulting apparent luminosity magnitude would be of the order of 15. 
 Now, such a bright and relatively nearby galaxy would be seen as extended object by {\it HST}/ACS.
  Indeed, the limiting magnitude for bright galaxies (above which they are observed as point sources) is around 23.3 (see \citealt{bono2010} and references therein; \citealt{blanton2009}). 
 No extended source has been observed in the field of \sou\ by {\it HST} and we can thus rule out the possibility of an extragalactic TDE.

\section{Conclusion}\label{sec:conclusion}
The nature of the puzzling new \integral\ source \sou\ is still unknown after a number of follow-up observations.  
We noted two empirical evidences typical of  TDEs:
 a $\propto t^{-5/3}$ trend of the light curve and  a thermal disc blackbody emission  almost constant with time.
 Thus, we started to investigate the possibility that a minor body was tidally disrupted by a compact object. 
 Our observational analysis and theoretical calculations indicate a free-floating terrestrial planet tidally disrupted by a WD.
In particular, the mass of the disrupted body is $M_{\rm mb} \approx 1.9 \times 10^{27} (M/M_{\rm Ch})$ g,
while the accreting object is consistent with a WD  close to the Chandrasekhar limit.

Finally, we have estimated the rate  of a planetary disruption event by a WD  in the GC NGC 6388:
although many of the relevant parameters are highly uncertain, 
we estimate the rate $\dot{N}_{\rm TE}$
to be in the range $\simeq 10^{-6} - 10^{-4} {\rm yr}^{-1}$. The lower limit suggests that the detection of such a event in the Milky Way is unlikely. 
However, if we consider the optimistic estimate for the rate, and considering  that the total number of GCs in the Galaxy is roughly 150, 
we obtain a total rate of events of about 0.05 yr$^{-1}$, i.e. one every $\sim 20$ yr, comparable with the lifetime of  \integral\ and \swift\ .

\section*{Acknowledgments}
We acknowledge interesting discussions with Giuseppe Bertin, Piergiorgio Casella, Domitilla De Martino, Alessandra De Rosa, Immacolata Donnarumma, 
Giuliana Fiorentino, Barbara Lanzoni, Cristina Pallanca, Patrizia Romano, Alexander Tschekoskoy, Dimitri Veras, Andrea Tiengo, Alice Zocchi.
MDS and GDC acknowledge financial support from the agreement
ASI-INAF I/009/10/0 and from PRIN-INAF 2009 (PI: L. Sidoli).
GL  acknowledges financial support from PRIN MIUR 2010-2011, {\it The
Chemical and Dynamical Evolution of the Milky Way and Local Group
Galaxies'}, prot. 2010LY5N2T.
MDS thanks the Mathematics and Physics Department  at the University of Salento for hospitality.
AN thanks INAF/IAPS for hospitality.

\label{lastpage}

\end{document}